\documentclass[12pt,a4paper]{article}
\usepackage{epsfig}
\usepackage[latin1] {inputenc}
\usepackage{amsmath} \usepackage{amsfonts} \usepackage{amssymb} \usepackage{graphicx} \usepackage{longtable}
\author{Les Hatton \\ leshatton.org\thanks{lesh@oakcomp.co.uk}}
\title{Power-laws and the Conservation of Information in discrete token systems:\\Part 2 The role of defect}
\begin{document}
\maketitle
\bibliographystyle{plain}
\begin{abstract}
In a matching paper \cite{Hatton2012a}, I proved that Conservation of Size and Information in a discrete token based system is overwhelmingly likely to lead to a power-law component size distribution with respect to the size of its unique alphabet.  This was substantiated to a very high level of significance using some 55 million lines of source code of mixed provenance.  The principle was also applied to show that average gene length should be constant in an animal kingdom where the same constraints appear to hold, the implication being that Conservation of Information plays a similar role in discrete token-based systems as the Conservation of Energy does in physical systems.

In this part 2, the role of defect will be explored and a functional behaviour for defect derived to be consistent with the power-law behaviour substantiated above.

This will be supported by further experimental data and the implications explored.
\end{abstract}
\begin{verbatim}
Keywords: Information content, defect clustering,
Component size distribution, Power-law
\end{verbatim}
\section{Preliminaries}
\subsection{Conservation of Information}
In \cite{Hatton2012a}, I showed that given a discrete token based system of M components where the $i^{th}$ component contains $t_{i}$ tokens, i = 1,..,M, conservation of a general quantity U and total Size T is overwhelmingly likely to lead to a size distribution which obeys

\begin{equation}
p_{i} = \frac{e^{-\beta \varepsilon_{i}}}{\sum_{i=1}^{M} e^{-\beta \varepsilon_{i}}}    \label{eq:varp}
\end{equation}

where the total size is given by

\begin{equation}
T = \sum_{i=1}^{M} t_{i}  \label{eq:con1}
\end{equation}

and there is some externally imposed entity $\varepsilon_{i}$ associated with each token of component \textit{i} whose total amount is given by

\begin{equation}
U = \sum_{i=1}^{M} t_{i} \varepsilon_{i} \label{eq:con2}
\end{equation}

I then showed that by identifying U with the total Hartley-Shannon information content \cite{Hartley1928}, \cite{Shannon1948}, \cite{Shannon1949}, \cite{Cherry1963}, that the resulting predicted distribution takes on a power-law distribution given by

\begin{equation}
p_{i} = \frac{(a_{i})^{- \beta}}{Q(\beta)} \sim (a_{i})^{- \beta}  \label{eq:pwrlaw}
\end{equation}

where $p_{i}$ is the probability of a component of size $t_{i}$ tokens occurring and $a_{i}$ is the size of the unique alphabet of tokens used to construct it.  Here

\begin{equation}
Q(\beta) = \sum_{i=1}^{M} e^{-\beta \frac{I_{i}}{t_{i}}}
\end{equation}

and the result is now subject to the twin constraints that the total number of tokens T is fixed

\begin{equation}
T = \sum_{i=1}^{M} t_{i}         \label{eq:T}
\end{equation} 

and the total Hartley / Shannon information content I, is also fixed

\begin{equation}
I = \sum_{i=1}^{M} I_{i}	\label{eq:I}
\end{equation}

where $I_{i}$ is the information content of the $i^{th}$ component given by

\begin{equation}
I_{i} = t_{i} log a_{i}     \label{eq:ii}
\end{equation} 

and log denotes the natual logarithm.
\paragraph{}
This result was substantiated against 55.5 million lines of source code in 6 languages and demonstrated valid to a p level $< 2.2.10^{-16}$.  In other words, it is extremely unlikely that this result would have occurred by chance.  (p levels $< 0.01$ are considered emphatic).

\subsection{Conservation of defect}
A defect in its simplest terms is a mistake.  In software systems, a defect is some kind of mistake in the coding, (and there are many kinds \cite{SoftTest}, \cite{SaferC}), which causes the run-time behaviour of a program to depart from its expected behaviour.  In a biological system, it might be a copying error in a gene.  In both cases, we can imagine that there must be a total number of defects D given by

\begin{equation}
D = \sum_{i=1}^{M} d_{i} \label{eq:con3}
\end{equation}

where $d_{i}$ is the number of defects in the $i^{th}$ software component or gene.

Following a similar development to \cite{HatTSE08}, this can be written as

\begin{equation}
D = \sum_{i=1}^{M} t_{i} (\frac{d_{i}}{t_{i}}) \label{eq:con4}
\end{equation}

If we now identify $\varepsilon_{i}$ of equation (\ref{eq:varp}) as follows,

\begin{equation}
\varepsilon_{i} = (\frac{d_{i}}{t_{i}})        \label{eq:vared}
\end{equation} 

(in other words, each token of the $i^{th}$ component has a defect density associated with it given by $(\frac{d_{i}}{t_{i}})$), then the corresponding most likely distribution \textit{which maintains constant total defects and size} is given from (\ref{eq:varp}) and (\ref{eq:vared}) by:-

\begin{equation}
p_{i} = \frac{e^{-\beta d_{i} / t_{i}}}{\sum_{i=1}^{M} e^{-\beta d_{i} / t_{i}}} \sim e^{-\beta d_{i} / \epsilon_{i}}   \label{eq:varpd}
\end{equation}

\textit{However, we know from the measurements described by \cite{Hatton2012a} that real software systems obey (\ref{eq:pwrlaw}) to a very high level of certainty} and so equating the distributions (\ref{eq:pwrlaw}) and (\ref{eq:varpd}) suggests that a defect conserving system will obey the following:-

\begin{equation}
d_{i} \sim t_{i} log a_{i}     \label{eq:tloga}
\end{equation} 

Note that this is an identical relationship to the information content of the $i^{th}$ component (\ref{eq:ii}) because they are both conserved during variation.

\subsection{Some existing component defect models}
It is interesting at this point to pause for a moment and consider empirically observed distributions of defects in components of real software systems.  The first thing to appreciate is that lines of code are inevitably used as a measure of program size in such studies.  The reason for this is that lines of code are \textit{much} easier to measure than the tokens used above, which require the development of compiler front-ends to measure properly, \cite{Hatton2012a}.  The downside however is that it is not a very precise measure in that lines of code can be defined in a number of ways, for example as a count of the newline characters as is most common, but it might also be a count of only those lines of code which cause a compiler to generate object code, (in which case they are known as executable lines of code).  In addition, lines are layout based and therefore subject to stylistic interpretation whereas tokens are unambiguous.  They are of course closely related but one programmer might typically use a smaller number of tokens per line than another as a matter of personal style.

The ease of use of lines of code as a measure has meant that virtually all of the research into empirical distributions of defect uses lines of code as the independent variable leading to a relationship , $d_{i} = d_{i}(n_{i})$, where $n_{i}$ is the number of lines of code in the $i^{th}$ component.

There have been numerous attempts at modelling such defect behaviour as a function of component size, for example, \cite{Aki71}, \cite{Lip82}, \cite{Gaf84}, \cite{Com90} and \cite{Hat97}.  In the absence of any models of defect growth, these are essentially exercises in data-fitting and all show at least linear growth in the number of defects with component size.  In particular, \cite{Lip82} and \cite{Hat97} both report logarithmic behaviour, and notably in the case of \cite{Lip82}, $d_{i} \sim n_{i} log n_{i}$.
\paragraph{}
\textit{In this section, I have shown that subject to the constraints of constant defect (\ref{eq:con3}) and constant size (\ref{eq:con1}), a component size model strongly substantiated in \cite{Hatton2012a} leads directly to a  prediction that the defect distribution for a software system in equilibrium will obey the relation (\ref{eq:tloga}).}
\paragraph{}
This is a direct consequence of the principle of Conservation of Information and this will now be tested on a two very disparate real systems which have been in use for some considerable time and should therefore exhibit at least quasi-equilibriated behaviour.  Note that in this sense, equilibriation refers to the process of continual use gradually flushing out residual defects so that as the number of discovered $\rightarrow$ D, (noting that we only know it is fixed, we do not know its value), the program becomes increasingly reliable or \textit{mature} as it is commonly known.

\section{Application to software systems}
\subsection{Experimental verification}
Validating the relationship (\ref{eq:pwrlaw}), although requiring the development of lexical analysers capable of extracting the required tokens \cite{Hatton2012a}, is unambiguous.  Such tokens are part of the definition of a programming language and when counted by separate experiments should yield the same results always, otherwise the language would have unacceptable ambiguity.

The situation is not so simple for the measurement of defect.  Such measurement almost always involves a measure of subjectivity, in the identification of the defect or even whether it is considered to be a defect at all.  Further complications intrude such as the counting of two code fragments in separate locations which together produce a defect.  Is this one occurrence or two ?  Such questions have never and probably will never be resolved unambiguously so it should immediately be recognised that defect measurements are noisy.  Token measurements are not, (unless the tokeniser itself is in error).

\subsection{Results}
With these comments in mind, two packages were initially selected to test the relationship in (\ref{eq:tloga}) because both have an extensive and well-maintained defect history which can be mined by suitable tools.

\subsubsection{NAG scientific subroutine library (Fortran)}
\begin{figure}
\centering
\begin{tabular}{cc}
\epsfig{file=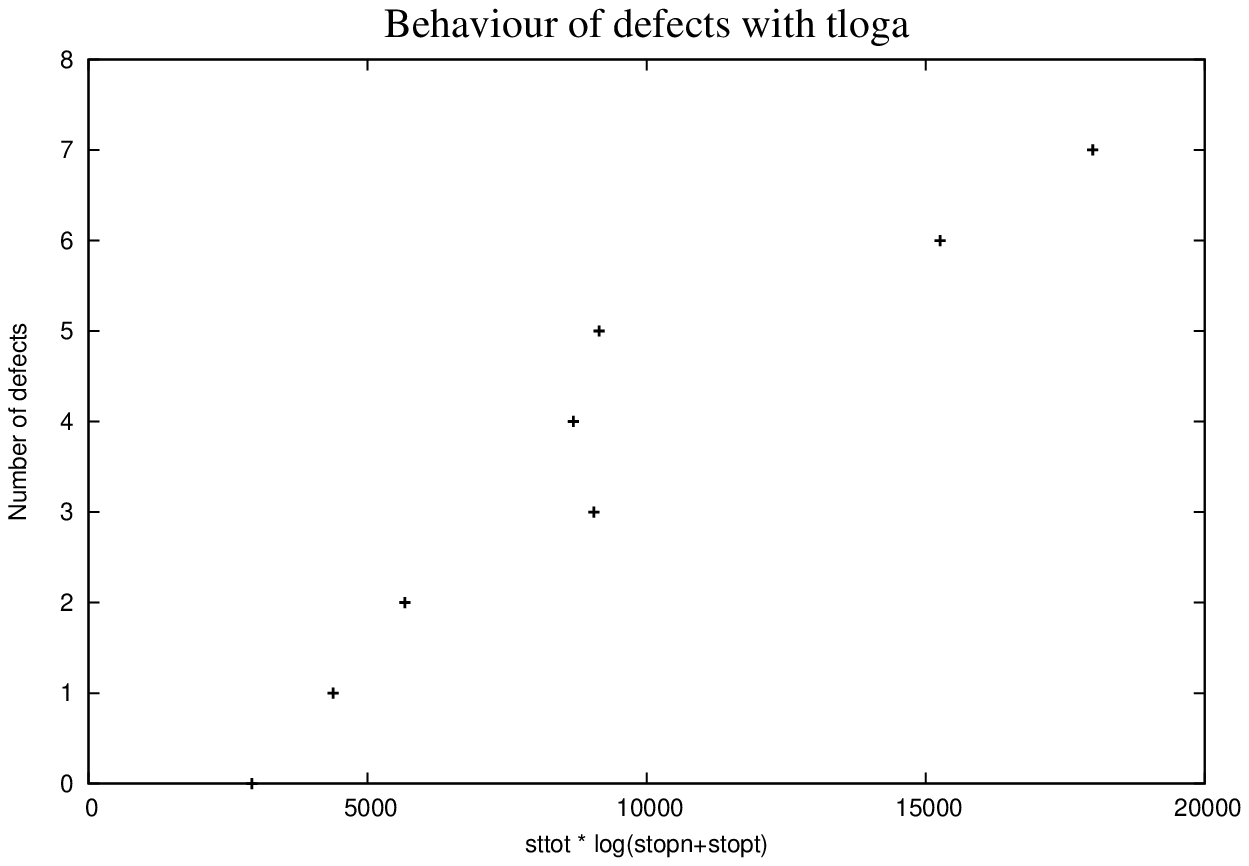,width=0.5\linewidth,clip=} &
\epsfig{file=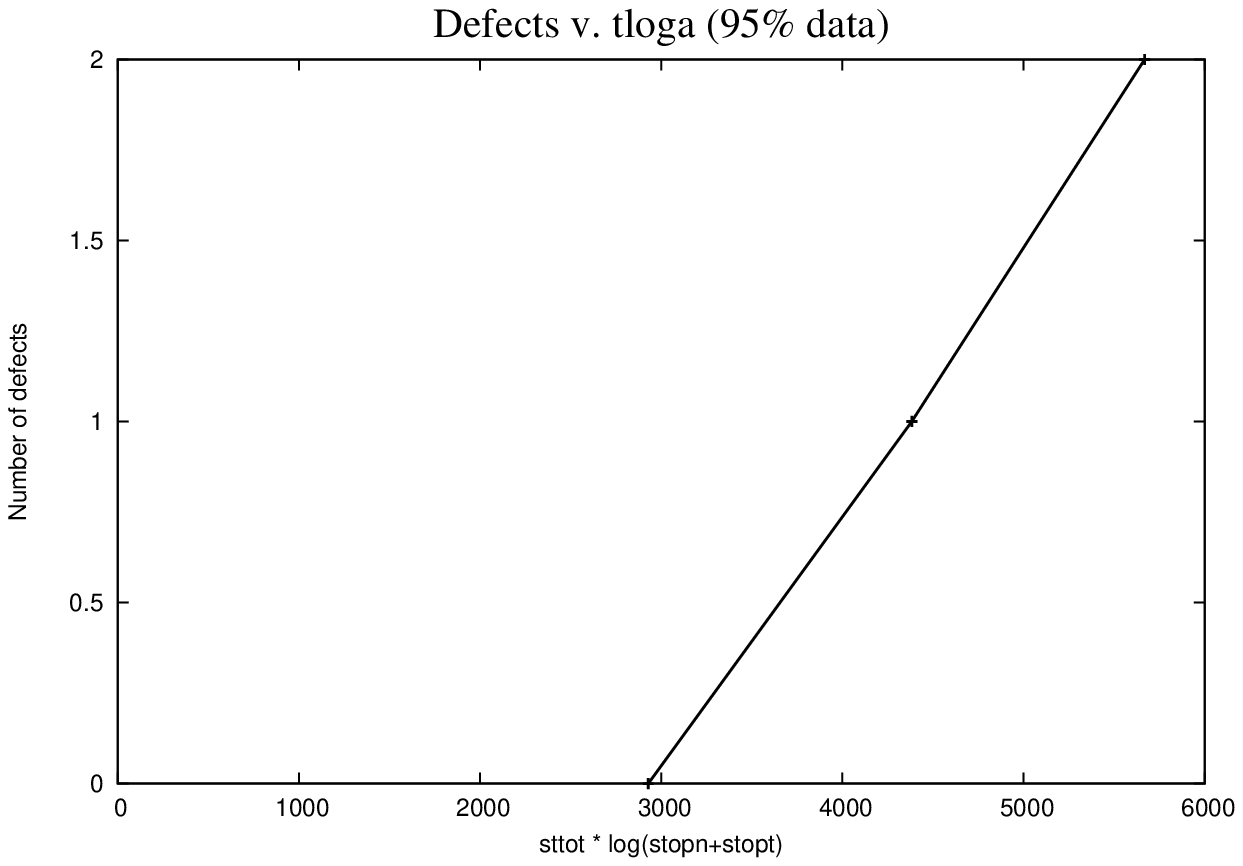,width=0.5\linewidth,clip=}
\end{tabular}
\caption{The distribution of defects (y-axis) against t log(a) (x-axis) for the NAG library.  The left hand graph shows all the data up to and including 7 defects.  The right hand graph shows those components with 0, 1 and 2 defects, (more than 95\% of the components).  It should be re-iterated that each point in the right hand graph is the mean of many values of tloga which have the same number of defects.}
\label{fig:tloga}
\end{figure}

The NAG Fortran scientific subroutine library was extensively analysed by \cite{HopHat08}.  It has a detailed defect record embedded in its source code which the authors mined and associated with each component so it can be merged with measurements of $t_{i}, a_{i}$ made on the same code.  For each defect up to a maximum of 7 per component, (very few components had more than this and were therefore excluded), the value of tloga was averaged and the resulting data is presented as Figure \ref{fig:tloga}.

The predicted linearity of Figure \ref{fig:tloga} was subjected to a standard test for significance using the linear modelling function lm() in the widely-used R statistical package\footnote{http://www.r-project.org/}.  This reported a high degree of linearity with an adjusted linear-fit correlation of 0.89, a high level of linear correlation with an associated p-value, (the probability of finding a dataset more unlikely than this one by chance) of 0.0002544, an emphatic result.  The corresponding output from R is shown below.  (Note that in the R analysis, the tloga values were normalised by a factor of 5000.0.)

\begin{verbatim}
lm(formula = y ~ x, data = universe)

Residuals:
    Min      1Q  Median      3Q     Max 
-0.7120 -0.4648 -0.3056  0.2195  1.4967 

Coefficients:
            Estimate Std. Error t value Pr(>|t|)    
(Intercept)  -0.6021     0.6048  -0.995 0.357931    
x             2.2439     0.2921   7.683 0.000254 ***
---

Residual standard error: 0.8036 on 6 degrees of freedom
Multiple R-squared: 0.9077,     Adjusted R-squared: 0.8924 
F-statistic: 59.03 on 1 and 6 DF,  p-value: 0.0002544
\end{verbatim} 

\subsubsection{Eclipse IDE (Java)}
The Eclipse IDE written in Java is another example of a well-instrumented software package.  In this case, the hard work of extracting defect records and associating them with particular components has already been done by \cite{NeuhausZimmermann2007}\footnote{See also http://www.st.cs.uni-sb.de/softevo. The data comes from releases 2.0,2.1 and 3.0. There are 10,613 components in the release 3.0}.  All that was necessary here was to extract all the $t_{i}, a_{i}$ using the methods described above and in \cite{Hatton2012a} and the data plotted for all components with up to 12 defects, (again very few components contained more than this and were consequently excluded).  Again the value of tloga was normalised by a convenient factor of 5000.0 before analysis with R.  The results this time were:-

\begin{verbatim}
lm(formula = y ~ x, data = universe)

Residuals:
    Min      1Q  Median      3Q     Max 
-1.9848 -0.6129 -0.2032  0.6618  1.7910 

Coefficients:
            Estimate Std. Error t value Pr(>|t|)    
(Intercept)   0.3256     0.5874   0.554     0.59    
x             1.5324     0.1340  11.435 1.91e-07 ***
---

Residual standard error: 1.133 on 11 degrees of freedom
Multiple R-squared: 0.9224,     Adjusted R-squared: 0.9153 
F-statistic: 130.8 on 1 and 11 DF,  p-value: 1.907e-07
\end{verbatim}

giving an adjusted R-squared of 0.92.  This represents an even high quality linear fit due to the larger quantity of data.

\begin{figure}
\centering
\begin{tabular}{cc}
\epsfig{file=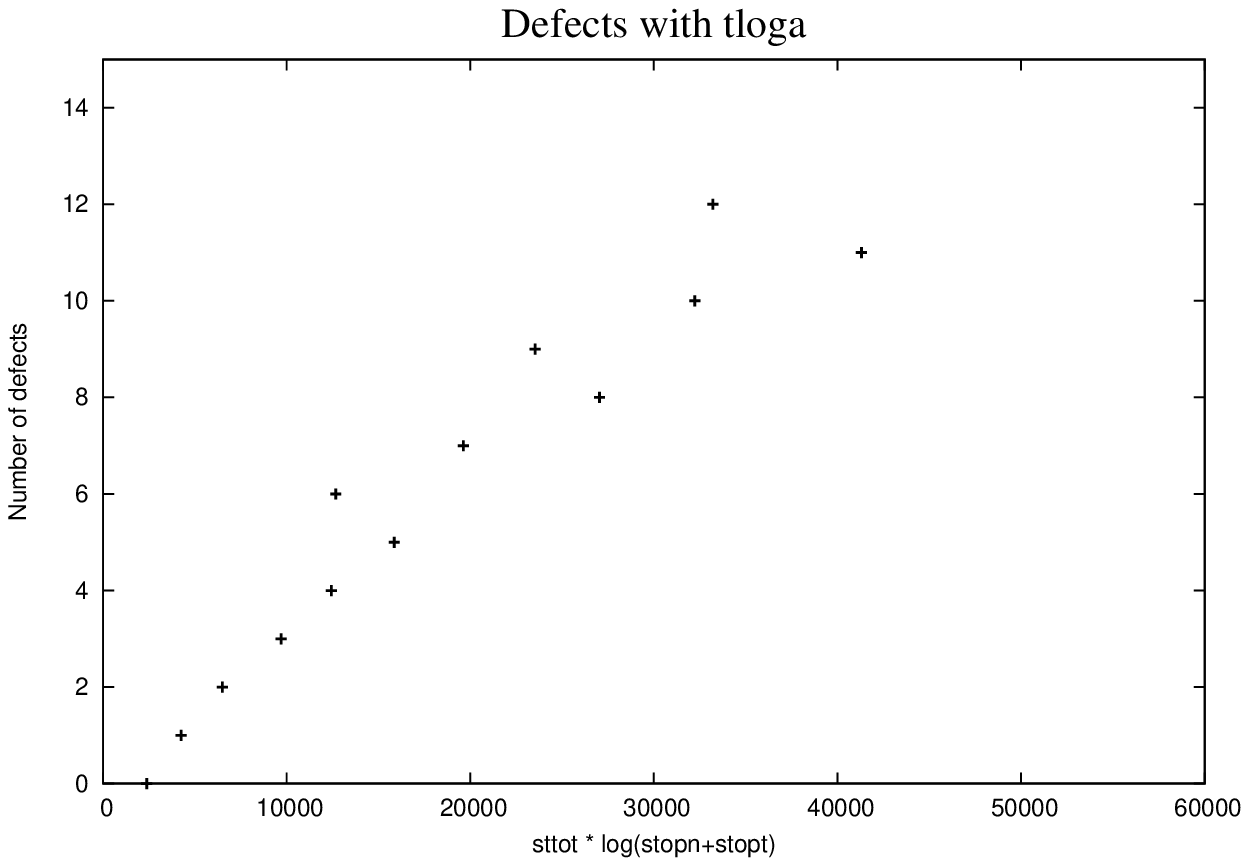,width=0.5\linewidth,clip=} &
\epsfig{file=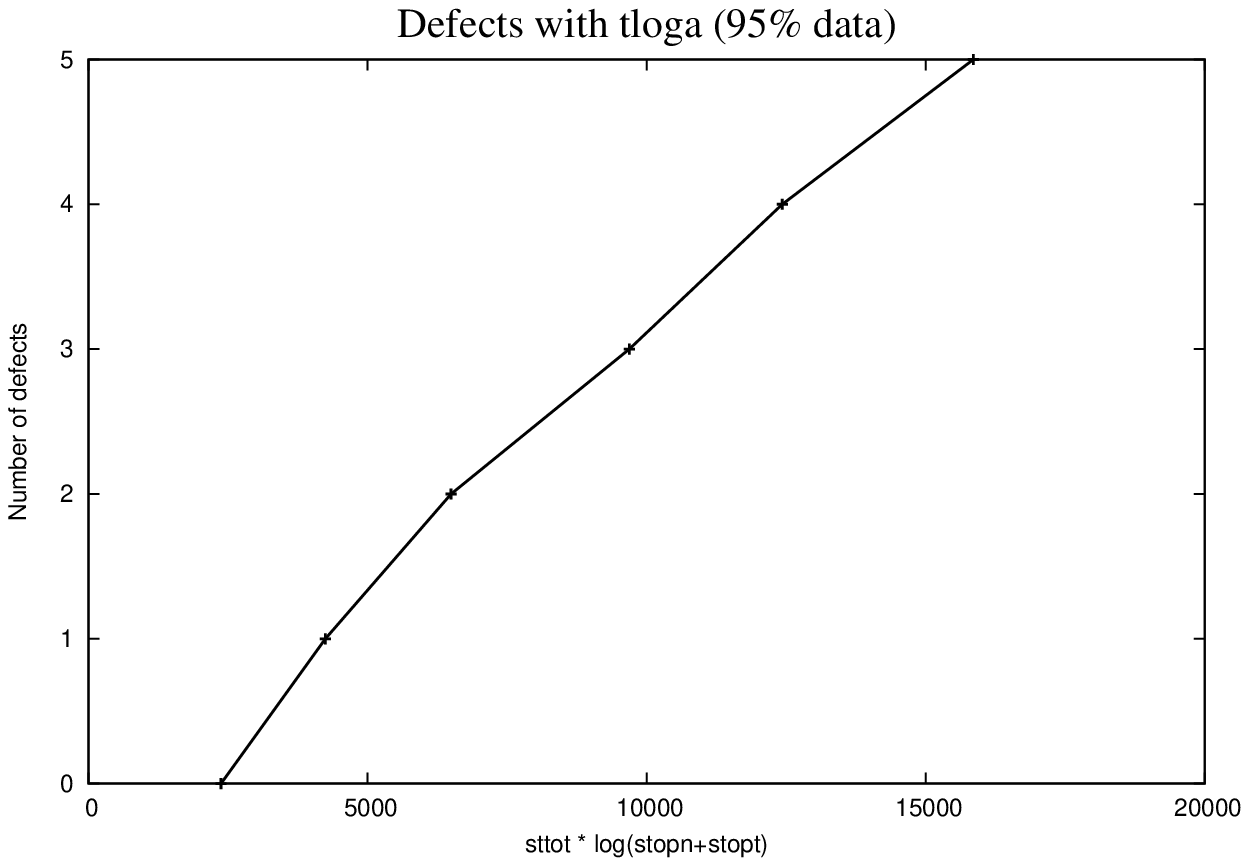,width=0.5\linewidth,clip=}
\end{tabular}
\caption{The distribution of defects (y-axis) against t log(a) (x-axis) for the Eclipse IDE library.  The left hand graph shows all the data up to and including 12 defects.  The right hand graph shows those components with 0-5 defects, (more than 95\% of the components).  Again each point in the right hand graph is the mean of many values of tloga which have the same number of defects.}
\label{fig:tloga_eclipse}
\end{figure}

\paragraph{}
To summarise these two experiments, even though defect data are inherently much noisier than token measurements, the degree of linearity predicted by (\ref{eq:tloga}) is well supported.

\subsection{Equilibriation}
This result may contribute to answering a difficult question in software engineering - ``How can you tell when a software component has been thoroughly tested ?''  This attempts to place into words the perceived property of a system which on continuous running in diverse environments, fails very rarely in some sense.  The problem is that when a product is shipped for the first time, a low early defect measure says nothing about the future behaviour unless it is linked with a substantial run-time history.  To put into the form of a simple aphorism:

\begin{quotation}
\textit{There are two ways of achieving low defect: the first is to have a very good system, and the second is to have very poor testing.}
\end{quotation} 

We obviously prefer the former.  However the development which led up to (\ref{eq:tloga}) considers its equilibriation as shown with a number of systems in \cite{Hatton2012a}.  In other words, departures from (\ref{eq:tloga}) may tell us how well the system has been tested.  It turns out that in the case of the NAG library, 95\% of the components have exhibited 0, 1 or 2 defects.  The same 95\% cut-off when applied to the Eclipse data covers 0 - 5 defects.    Inspecting Figures \ref{fig:tloga} and \ref{fig:tloga_eclipse} for these values shows that they are very highly linear here with significant departures only appearing for a small population of components.  If this 5\% of components are excluded in both cases, the adjusted R-squared values reach 0.99 in both cases.

\begin{quotation}
\textit{I therefore propose that the adjusted R-squared value for linear fit could be used to determine how well code has been tested simply from its defect records.  If there is no substantial evidence of linearity up to a number of defects corresponding to say, 95\% of the defect data, then the defect data has not yet equilibriated and it is likely that there are more defects to be found.  This is a form of reliability growth modelling in which the temporal axis of reliability growth is replaced by departures from linearity of an asymptotic defect distribution shaped by the Conservation of Information.}
\end{quotation}

Software defect data are not easy to work with as has already been discussed but it is hoped that this will inspire further experiments to test the asymptotic result of (\ref{eq:tloga}).

\section{Application to genetic systems}
I will now apply the general principle expressed by (\ref{eq:tloga}) to predicting defect properties of genes.  No experimental evidence will be presented for this here as this is the subject of a companion paper, \cite{HattonWarr2012}.

\begin{figure}
\includegraphics[width=10cm,height=8cm]{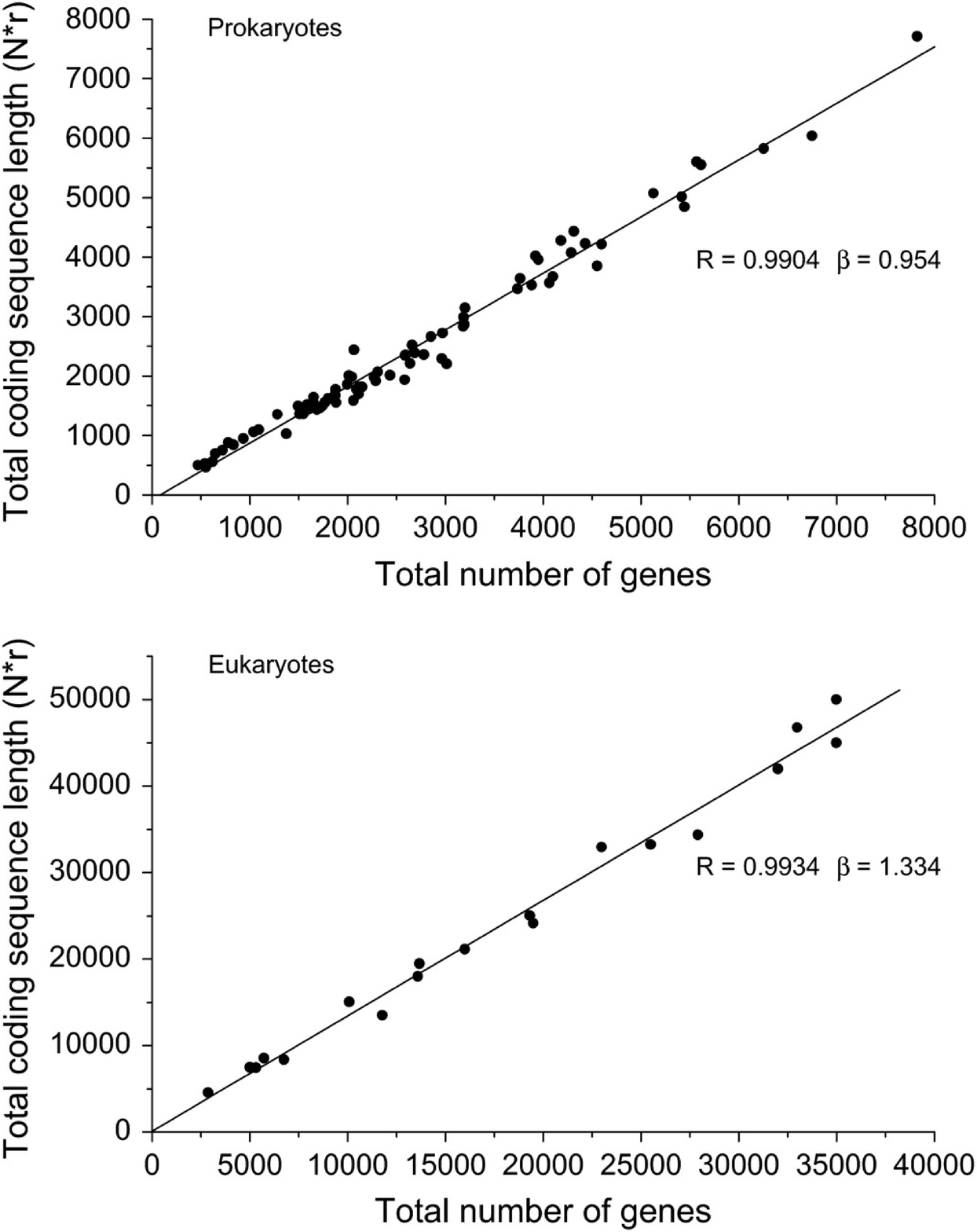}
\caption{Linear regression analysis of the total sequencing length against the number of genes shown by \cite{XuJune2006}.}
\label{fig:xu2006}
\end{figure}

In \cite{Hatton2012a}, I demonstrated that the principle of Conservation of Information predicts that gene length is uniformly distributed, a direct result of the fixed 4-base \textit{ACGT} alphabet of the genome.  In turn this implies that the ratio of total sequence length / number of genes is constant.  This prediction is well supported by the experiments of \cite{XuJune2006} and for continuity, I will repeat some of their comments here.  They surveyed almost all prokaryotic and eukaryotic species whose complete genome sequence data were then available and well annotated. These data included 81 prokaryotes and regressed the estimate of total coding sequence length against the estimate of the number of genes for each of the two groups of species.  They found that although the average lengths of genes in prokaryotes and eukaryotes are significantly different, \textit{the average lengths of genes are highly conserved within either of the two kingdoms}. They concluded that natural selection has clearly set a strong limitation on gene elongation within the kingdom and that the average gene size adds another distinct characteristic for the discrimination between the two kingdoms of organisms.  Their data is reproduced by kind permission as Figure \ref{fig:xu2006}.

\subsection{The growth of defects on genes}
As noted in \cite{Hatton2012a}, genes have length $t_{i}$ bases chosen from a unique alphabet of $a_{i}$ bases, however the alphabet of bases in genetic codes is \textit{fixed} to adenine, cytosine, guanine and thiamine.  In other words, $a_{i} = 4, \forall i$.  Using (\ref{eq:tloga}) for the genome then gives

\begin{equation}
 d_{i} \propto t_{i}      \label{eq:dvt}
\end{equation}

In other words, Conservation of defect in a system with uniform probability distribution for gene length implies that by far the most likely outcome is that genetic defects are linearly proportional to gene length.

This is considered in much more detail including the effects of kingdoms in a companion paper \cite{HattonWarr2012}.

\section{Comparisons with Halstead}
This is not the first time in which defects have been related to tokens of programming languages.  Halstead \cite{Halstead75}, \cite{Halstead77} made an intensive study of the relationship between Shannon information theory and programming language defining a number of heuristics, program volume, effort, information content and so on.  The current work is based on \cite{HatTSE08}, \cite{Hatton2011a}, and takes a different approach combining Shannon information theory with concepts of statistical mechanics.  This avoids emphasising the meaning of tokens and simply refers to the choices which can be made as described by \cite{Cherry1963}.

\section{On defect growth generally}
Defect growth in systems has been widely studied for a number of years using a variety of reliability growth models, \cite{Brocklehurst1992}, \cite{Musa1993}, \cite{Barghout1997}, \cite{Littlewood2000}.  In spite of this, it is still relatively rare to find good defect data which can be analysed for the verification of models such as that proposed here.  The Open Source movement has improved this along with tools such as Bugzilla\footnote{http://www.bugzilla.org/} but the situation is still not as good as for the analysis of token distributions in open source in part 1 of this paper, \cite{Hatton2012a}.  In particular, equilibriation to the predicted distribution (\ref{eq:tloga}) is not well covered as it requires meticulous defect records from the early days of a large system and these need to be associated with particular components as done in an exemplary fashion by \cite{NeuhausZimmermann2007} with Eclipse.

\section{On equilibriation and tokens}
Perhaps the most difficult idea to grasp in using variational principles like this is that such principles are ergodic.  They are not talking about a single system but about all possible systems.  In other words, when total size is constrained to say T tokens, this does not mean that the results are only relevant to systems with this size.  Instead, all the variational method says is that if the totality of all possible systems of size T are considered, then an overwhelmingly large number of them will produce a component size distribution obeying (\ref{eq:pwrlaw}).

If I select a particular system and change its size in some way to T' as occurs in both software development, through incremental change and also in genetic development, through the usual mechanisms of natural selection and mutation, then it simply becomes one of the totality of systems of size T'.  The variational method knows nothing of this and indeed doesn't care.  I could, because I have free choice, develop a software system of size T with M components in the programming language C all but one of which contains the same static function definition along with an empty main() component.  It will compile, link, run and be exceptionally uninteresting in every way except that it will not obey a power-law in its component distribution.  It is however, just one of the totality of programs of size T, which overwhelmingly will obey such a power-law.

I could embark on a crusade and try to persuade every programmer on earth to write the same program for the rest of their lives and to pass this on to their descendants in order to break the power-law distribution but it is not very likely and in any case, ergodically speaking, it does not cover every system of T tokens.  Finally, in a perfect gas, which is where such variational methods were honed, it is perfectly possible that all the molecules in a room will suddenly find themselves under a table so it shoots into the air, but it is not very likely.
\paragraph{}
It is also worth saying something about tokens in general and unique alphabets of tokens in particular.  When deciding on a unique alphabet in a programming language, it is easy to find token combinations which are dependent on each other.  For example, again from the programming language C, the token ``if'' \textit{must} be followed by the token ``(''.  Anything else is syntactically illegal in C.  Does this then mean that these are one token or two ?  The answer to this conundrum lies with information theory.  As I discussed at some length in \cite{Hatton2012a}, the meaning of the tokens is \textit{irrelevant} in this context.  Information content is only about choice, not meaning, so there are indeed two tokens.

\section{Conclusions}
The paper presents several contributions.
\begin{itemize}
\item Using variational principles suggested in \cite{HatTSE08}, \cite{Hatton2011a} and using the principle of the Conservation of Information, it is predicted that the number of defects in any component of size $t_{i}$ tokens constructed from a unique alphabet of $a_{i}$ tokens, will equilibriate to a distribution given by,
\begin{equation}
 d_{i} \sim t_{i} . log(a_{i})     \label{eq:concl1}
\end{equation}
Substantial evidence in favour of this was presented using a large Fortran system, (the NAG library) and a Java system, (the Eclipse IDE).  Note that this distribution has clustering properties.  Observation of defect clustering is described in \cite{Hatton2012c}.
\item It is proposed that departures from (\ref{eq:concl1}) could be used to measure the degree of equilibriation which has taken place, specifically the adjusted R-value of a linear fit.
\item The underlying principle of Conservation of Information and constant total number of defects lead to a prediction that the number of defects in a gene is linearly proportional to its length.  This raises a number of issues and is being considered separately in a companion paper \cite{HattonWarr2012}. 
\end{itemize}

\section{Acknowledgements}
I am grateful to Professor Peter Bishop of City University for pointing out the relationship of Halstead's work to my own.  I would also like to thank Professor Zewei Luo
 of the University of Birmingham for his kind permission to reproduce the important result of Figure \ref{fig:xu2006}, as far as I know, the only such data on this intriguing subject.  The Eclipse defect data were carefully extracted and made available by Professor Andreas Zeller and his team at Saarland University.

\bibliography{lh_biblio}
\end{document}